\documentclass{article}

\PassOptionsToPackage{numbers, compress}{natbib}

\usepackage[final]{neurips_2023}

\usepackage[utf8]{inputenc} %
\usepackage[T1]{fontenc}    %
\usepackage{hyperref}       %
\usepackage{url}            %
\usepackage{booktabs}       %
\usepackage{amsfonts}       %
\usepackage{nicefrac}       %
\usepackage{microtype}      %
\usepackage{xcolor}         %

\usepackage{soul}
\newcommand{\ours}{our method}
\newcommand{\Ours}{Our method}
\newcommand{\absolut}{Absolut!\ }

\newcommand{\figref}[1]{Fig.~\ref{#1}}
\newcommand{\tabref}[1]{Table~\ref{#1}}
\newcommand{\secref}[1]{Section~\ref{#1}}
\newcommand{\algref}[1]{Algorithm~\ref{#1}}
\usepackage{glossaries}
\newacronym{rl}{RL}{Reinforcement Learning}
\newacronym{ai}{AI}{artificial intelligence}
\newacronym{mdp}{MDP}{Markov Decision Process}
\newacronym{sql}{SQL}{Structured Q-learning}
\newacronym{rr}{RR}{Replay Ratio}
\newacronym{auc}{AUC}{Area under the Curve}
\newacronym[plural=AAs,firstplural=amino acids (AAs)]{aa}{AA}{amino acid}
\newacronym{dqn}{DQN}{Deep Q-networks}
\usepackage{graphicx}
\usepackage{algorithm}
\usepackage{algorithmic}
\usepackage{amssymb}
\DeclareMathOperator*{\argmax}{arg\,max}
\DeclareMathOperator*{\maxmin}{Max\,min}

\title{Stable Online and Offline Reinforcement Learning for Antibody CDRH3 Design}

\author{%
  Yannick Vogt$^{1,2,}$\thanks{Corresponding author: vogty@cs.uni-freiburg.de} , Mehdi Naouar$^{1,2}$, Maria Kalweit$^{1,2}$, Christoph Cornelius Miething$^{2,3}$, \\
  \textbf{Justus Duyster$^{3}$, Roland Mertelsmann$^{2,3}$, Gabriel Kalweit$^{1,2,}$\thanks{These authors contributed equally.} ~, Joschka Boedecker$^{1,2,4,}$\footnotemark[\value{footnote}]}
  \\
$^1$University of Freiburg, $^2$Collaborative Research Institute Intelligent Oncology (CRIION), \\ $^3$University Medical Center Freiburg, $^4$BrainLinks-BrainTools
  }

\begin{document}

\maketitle

\begin{abstract}The field of antibody-based therapeutics has grown significantly in recent years, with targeted antibodies emerging as a potentially effective approach to personalized therapies.
Such therapies could be particularly beneficial for complex, highly individual diseases such as cancer.
However, progress in this field is often constrained by the extensive search space of amino acid sequences that form the foundation of antibody design.
In this study, we introduce a novel reinforcement learning method specifically tailored to address the unique challenges of this domain.
We demonstrate that our method can learn the design of high-affinity antibodies against multiple targets \textit{in silico}, utilizing either online interaction or offline datasets.
To the best of our knowledge, our approach is the first of its kind and outperforms existing methods on all tested antigens in the \absolut database.
\end{abstract}

\section{Introduction}
\label{sec:introduction}
Antibody-based therapeutics have revolutionized the treatment of various conditions~\citep{norman2019computational, robert2022unconstrained, kaplon2023antibodies}. Especially for highly individual diseases like cancer, antibody-therapeutics offer great potential in guiding the cellular immune response. However, the process of designing \textit{de novo} antibodies that bind effectively to given targets poses substantial challenges, primarily due to the combinatorial explosion of potential amino acid sequences that need to be considered. Research has shown that the CDRH3 region's diversity within an antibody suffices for most antibody specificity~\citep{xu2000diversity}. Consequently, designing CDRH3 sequences is typically used as a surrogate for the design of entire antibodies, as this reduces the search space significantly. Still, the search space for CDRH3 regions consisting of a combination of twenty naturally occurring \glspl*{aa} grows exponentially with $20^L$, where $L$ is the length of the CDRH3 regions. 

In recent years, \gls*{rl} methods~\citep{sutton1998reinforcement} have demonstrated their efficacy in hard exploration tasks and search spaces with high branching factors, as exemplified by AlphaGO~\citep{silver2017mastering}, and learning from pre-collected datasets~\citep{levine2020offline}. In addition, transformer-based language models like GPT~\citep{radford2019language, brown2020language} have exhibited remarkable advancements in the domain of natural language processing. Leveraging the analogy between natural language and protein amino acid sequences, these models have proven valuable for protein-related tasks~\citep{elnaggar2023ankh}. Taking up on these two recent developments, our method combines \gls*{rl} with transformer-based neural networks to effectively design binding antibody CDRH3 sequences \textit{in silico}. 

Applying \gls*{rl} in this domain, however, appears challenging, as current sophisticated methods typically rely on huge datasets acquired through an agent's active engagement and direct interaction with its environment. Consequently, a method that can learn from pre-collected datasets would offer a great advantage in allowing for consecutive reuse of previous trials. This is why we place particular emphasis on the \emph{offline} setting. Offline \gls*{rl} in turn often faces challenges related to overestimation and stability~\citep{levine2020offline}. Combining insights from \citet{lan2020maxmin}, \citet{an2021uncertainty}, and \citet{ball2023efficient}, we tackle overestimation by training an ensemble of Q-functions employing LayerNorm to approximate the binding energy of a given \gls*{aa} sequence. To address the issue of the high branching factor, we additionally employ an efficient implementation for \emph{oversampling} of good candidates and shape the immediate reward to make small differences more apparent.

Our agent aims at minimizing the binding energy or maximizing the \emph{affinity} between candidate antibody and target antigen, representing their strength of interaction, as this is closely tied to its therapeutic efficacy~\citep{makowski2022co}. The binding energy in turn depends on many physical factors emerging from the structure and binding locations of candidate and target. To evaluate the affinity of antibody-antigen complexes \textit{in silico} efficiently, we rely on the \absolut software~\citep{robert2022unconstrained}. Complexes are thereby discretized onto a lattice representation using their alpha-carbon locations since current methods utilizing continuous structure representations have shown to be error-prone or too time-consuming~\citep{buel2022can, lin2023evolutionary}.

Two methods benchmarked on \absolut are \gls*{sql}~\citep{cowen_rivers2022structured} and AntBO~\citep{khan2022antbo}. The first shares similarities with our approach as it utilizes \gls*{rl} principles and employs an attention-based neural network. However, the authors observed that designing sequences step-by-step resulted in suboptimal binders, whereas generating the sequence as a whole improved performance. In contrast, our method adopts a step-by-step approach allowing the agent to take incomplete sequences into account, resulting in even better binders, and is directly applicable to the offline setting. The second is based on Bayesian Optimization using Gaussian processes, demonstrating high sample efficiency. However, the cubic complexity of Gaussian processes in relation to the number of samples limits the scalability of such approaches.

Our contributions are threefold. First, we introduce the (to the best of our knowledge) first offline \gls*{rl} method for antibody design, moving towards real-world applicability in this domain. Second, we show that our method is applicable to both the online and offline settings without any modification. Third, we introduce a new state-of-the-art on all tested antigens in the \absolut benchmark.

\section{Background}
In this section, we briefly introduce the necessary background for our method and experiments.
\label{Background}
\paragraph{Markov Decision Process}
In the following, we describe how to model the task of designing a CDRH3 region as a \gls*{mdp}.

Generally, we consider an \gls*{mdp} for deterministic fixed horizon systems as a tuple $\langle S, A, T, R, s_0\rangle$, where $S$ is the set of states, $A$ is the set of actions, and $T:S\times A \rightarrow S$ is the deterministic transition function, where $T(s,a)$ represents the state that the MDP will transition to from state $s$ under action $a$. $R:S\times A\rightarrow\mathbb{R}$ is a reward function, where $R(s,a)$ represents the immediate reward $r$ received after taking action $a$ in state $s$. The initial state of the MDP is represented by $s_0\in S$.

In our setting, the initial state $s_0$ is given as a sequence of blank tokens $[[B_{i}]_{i=0}^{L-1}]$, representing the absence of any \glspl*{aa}, where $L$ is the desired length of the sequence.
The \absolut software, used to evaluate designed CDRH3 sequences, limits the length of a CDRH3 to $L=11$ \glspl*{aa}.
All other states are then derived from the initial state by replacing the tokens starting from $B_0$ to $B_{10}$. The transition function, therefore, transforms a state $s_t$ to a next state $s_{t+1}$ where $B_t$ is replaced by the current action $a_t \in A$ where $A$ is the set of twenty naturally occurring \glspl*{aa}, resulting in $[[a_{i}]_{i=0}^{t}, [B_j]_{j=t+1}^{L-1}]$.
The reward function $R$ returns zero for all states and actions except for the $a_{10}$ which completes the CDRH3 sequence. There, the reward is derived from the binding energy computed by the \absolut software for the designed CDRH3 and the given target antigen.

\paragraph{Online and Offline Reinforcement Learning}
The goal of an \gls*{rl} agent interacting with an MDP is to learn a policy $\pi:S\rightarrow A$, which maps each state to an action, that maximizes the expected sum of future rewards until a fixed terminal time step $L-1$, defined as $\sum_{t=0}^{L-1}R(s_t, a_t),$ where $s_t$ is the state at time $t$ and $a_t$ is the action taken at time $t$ under policy $\pi$. 
We approximate this sum using a Q-function $Q:S\times A\rightarrow \mathbb{R}$ represented by a neural network. Consequently, the policy $\pi$ is derived from $Q$ as $\pi(s) = \argmax_a Q(s,a)$.

The field of \gls*{rl} divides into two popular sub-paradigms: \emph{online} and \emph{offline} \gls*{rl}.
The more common \emph{online} \gls*{rl} describes agents learning from self-executed actions in the environment, whereas \emph{offline} methods describe agents learning solely from pre-existing datasets. The latter gained more and more traction in recent years, especially in domains where data collection can be tedious or dangerous~\citep{levine2020offline}. We, therefore, consider the offline setting more feasible for antibody design, where the exploitation of growing and older datasets instead of time-intensive online evaluations appears significant and could prove highly valuable.
In both paradigms and our domain model-based \gls*{rl} can be considered~\citep{angerm_uller2020model, jain2022biological}. We refrain from model usage to omit possible exploitation or bias of the learned model~\citep{trabucco2021conservative}.
\section{Stable Reinforcement Learning for Antibody Design}
\label{method}
In the following, we explain the components of our off-policy \gls*{rl} algorithm. Further, we visualize \ours{}{} and its interaction with the \absolut software in \figref{cover}.
We provide pseudocode, the full network architecture, and additional details in the Appendix.
\begin{figure*}[t]
\begin{center}
\centerline{\includegraphics[width=0.7\textwidth]{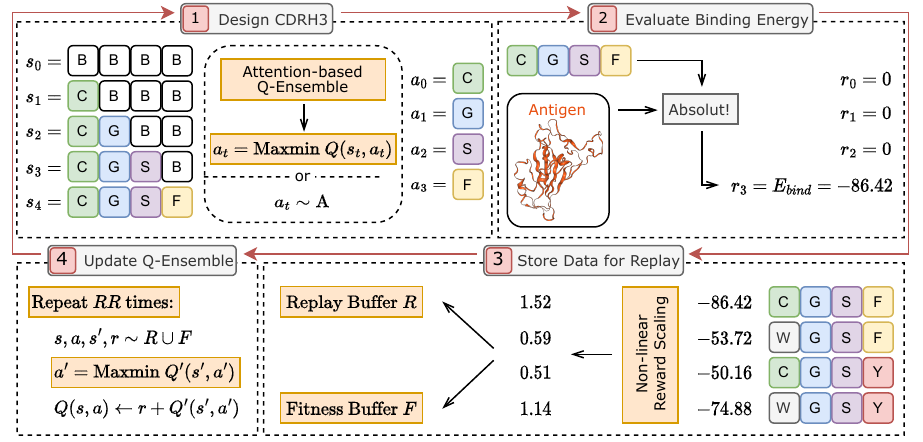}}
\caption{Visualization of \ours{}{} on fictive CDRH3 sequences of length four. \Ours{} repeatedly designs CDRH3 sequences (1), and evaluates them using the \absolut software (2), stores the gathered data (3), and updates its Q-function and thereby policy (4). We highlight important components of \ours{} in orange.}
\label{cover}
\end{center}
\end{figure*}
\paragraph{Favor Exploration -- Replay the Fittest}
In proteins, the effect of a mutation depends on the protein sequence as a whole. Therefore, two changes in the sequence which, isolated, each reduce the protein's utility can have a synergistic positive effect when they occur at the same time. This effect, which is called epistasis, can prevent proteins from improving ~\citep{wu2016adaptation}. We hypothesize that a similar effect may be prevalent when an antibody-designing \gls*{rl} agent emphasizes exploitation over exploration, resulting in only single changes.
We thus advocate for a high number of random actions sampled from the action space, $a\in A$, during training.

To counterbalance reduced exploitation and to over-sample the very best sequences during replay, we introduce a \emph{Fitness Buffer}.
Implemented as a min-heap, this buffer contains transitions corresponding to the $k$ best sequences discovered. Intuitively, this additional buffer facilitates learning from infrequently occurring sequences.
During replay, the agent is trained on a mixture of samples from the Fitness Buffer and a conventional Replay Buffer which stores all observed transitions (cf. \algref{alg:summary} in the Appendix). 
This technique can be seen as an adaption of rank-based Prioritized Experience Replay~\citep{schaul2016prioritized} where the importance of a transition corresponding to one of the best sequences is multiple times higher than that of a common sequence.

\paragraph{Stabilizing Reinforcement Learning with Ensembles}
When dealing with large search spaces, such as protein sequences, it is infeasible to explore each possibility. Therefore, there will always be actions that are out of the training distribution of the \gls*{rl} agent for a given state. The value of such actions is often erroneously overestimated~\citep{hasselt2010double,levine2020offline}.
We approach this issue by utilizing an ensemble of $N$ learned Q-functions as in Maxmin Q-learning~\citep{lan2020maxmin} and many offline \gls*{rl} algorithms~\citep{an2021uncertainty}, and define the Maxmin-policy as $\maxmin Q(s,a) := \argmax_{a\in A}\min_{i\in\{1, ...,N\}}Q_i(s, a)$. We hypothesize that this design choice allows us to apply the agent to both online and offline \gls*{rl}.

\paragraph{Attention-based Q-Networks}
Similar to \gls*{sql} and large language models~\citep{radford2019language}, which can process natural language, we use a Q-network based on multi-head self-attention to process a sequence of amino acids. To facilitate the training of such an architecture we propose using a high \gls*{rr}~\citep{d_oro2023sample}, meaning many updates to the network per environment interaction.

\paragraph{Reward Scaling for Binding Energy}
To amplify the difference in binding energy values, we non-linearly scale the reward to $r_{scaled} = (\frac{r}{\beta})^2$.
Thereby, $\beta$ is a constant to reduce the magnitude of the rewards.
This amplifies both the difference between weak binders and strong binders as well as the difference between strong binders where binding energy values may differ only minimally, often down to mere fractions of a unit. We, therefore, argue that this modulation of the immediate reward facilitates the noisy learning progress of neural networks in \gls*{rl}.

\section{Results and Discussion}
\label{results}
In this section, we present and discuss the results of our experiments both in the online and offline settings. To enable a comparison, we evaluate \ours{} on the same antigens as \gls*{sql} and AntBO. The hyperparameter settings, as well as an analysis of statistical significance and an ablation study, are given in the Appendix.

\begin{table*}[t]
\caption{Final binding energy per method and antigen. Shown is \ours{} in the online and offline setting, as well as SQL and AntBO. We mark results significantly outperforming AntBO with $^{1}$ and results significantly outperforming both AntBO and SQL with $^{2}$. \absolut and Online Lead represent the best energy contained in the \absolut database and offline dataset (collected using an online agent) respectively. Mean and standard deviation are given over eight and ten seeds in \ours{} and \gls*{sql} respectively.}
\label{large-table}
\begin{center}
\begin{footnotesize}
    \begin{tabular}{lcccccc}\toprule
    Antigen & Ours Online & SQL & AntBO & Absolut! & Ours Offline & Online Lead \\ 
    \midrule
    3RAJ\_A & -123.09 $\pm$ 1.74$^{1}$& -123.00 $\pm$ 2.88  & -119.07 & -116.74 & -125.08 $\pm$ 0.09$^{2}$& -125.19 \\ 
    2DD8\_S & -131.06 $\pm$ 1.36$^{1}$& -129.00 $\pm$ 4.10  & -127.57 & -125.07 & -132.37 $\pm$ 0.11$^{1}$& -132.45 \\ 
    1ADQ\_A & -112.55 $\pm$ 0.06$^{2}$  & -112.00 $\pm$ 0.41  & -110.70 & -108.53 & -112.56 $\pm$ 0.06$^{2}$ & -112.58 \\ 
    1OB1\_C & -111.27 $\pm$ 0.98$^{1}$& -111.00 $\pm$ 0.51  & -108.78 & -108.78 & -111.78 $\pm$ 0.08$^{2}$ & -111.88 \\ 
    2YPV\_A & -116.48 $\pm$ 0.91$^{1}$ & -115.00 $\pm$ 2.36  & -108.75 & -114.47 & -117.30 $\pm$ 0.08$^{2}$ & -117.37 \\ 
    1NSN\_S & -107.34 $\pm$ 0.35$^{2}$& -106.00 $\pm$ 1.70  & -105.02 & -107.16 & -107.65 $\pm$ 0.06$^{2}$ & -107.75 \\ 
    1WEJ\_F & -88.66 $\pm$ 0.58$^{1}$& -88.30 $\pm$ 0.13  & -86.89 & -86.03 & -89.99 $\pm$ 0.23$^{2}$ & -90.11 \\ 
    2JEL\_P & -88.15 $\pm$ 1.44$^{2}$& -84.70 $\pm$ 1.62  & -84.33 & -86.05 & -90.54 $\pm$ 0.03$^{2}$& -90.57 \\ 
    \bottomrule
    \end{tabular}
\end{footnotesize}
\end{center}
\end{table*}

\begin{figure}[h!]
\begin{center}
\begin{minipage}[t]{0.49\textwidth}
\begin{center}
\centerline{\includegraphics[width=0.9\textwidth]{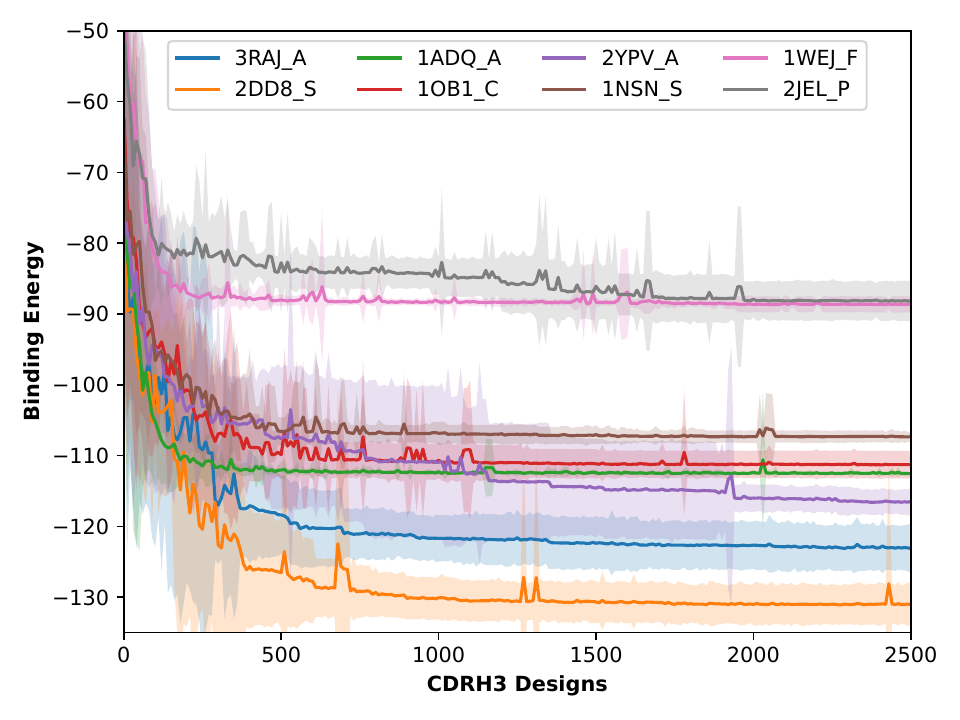}}
\end{center}
\end{minipage}
\begin{minipage}[t]{0.49\textwidth}
\begin{center}
\centerline{\includegraphics[width=0.9\textwidth]{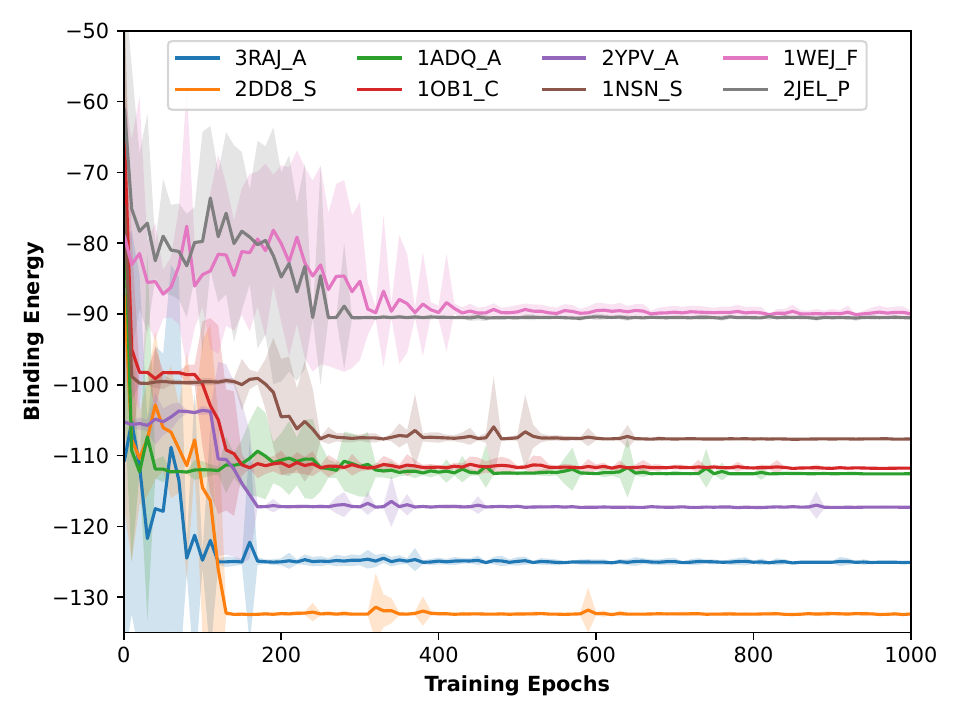}}
\end{center}
\end{minipage}
\end{center}

\caption{Progress of the binding energy of our method on different antigens, colored as displayed in the provided legend. The left plot depicts the online setting, while the offline setting is shown in the right plot. Mean and two standard deviations over eight seeds are shown.}
\label{convergenceplot}
\label{progress_offline}
\end{figure}

\paragraph{Online Learning}
\label{online}
In \tabref{large-table} we report the binding energy reached with our method, as well as \gls*{sql}~\citep{cowen_rivers2022structured}, and AntBO~\citep{khan2022antbo}. 
Further, we report the lowest binding energy of 6.9 million murine CDRH3 sequences in the \absolut database.
We observe that \ours{} outperforms or matches binding energies found by existing methods on all tested antigens and exceeds the best values in the \absolut database.
Note, that we use more CDRH3 design steps during our experiments than \gls*{sql} and AntBO. In \figref{convergenceplot} we visualize the evaluation progress of our method on different antigens. Notably, the evaluation progress is stable and does not collapse.

\begin{figure}[h]
\begin{minipage}[t]{0.49\textwidth}
\begin{center}
\centerline{\includegraphics[width=1.0\textwidth]{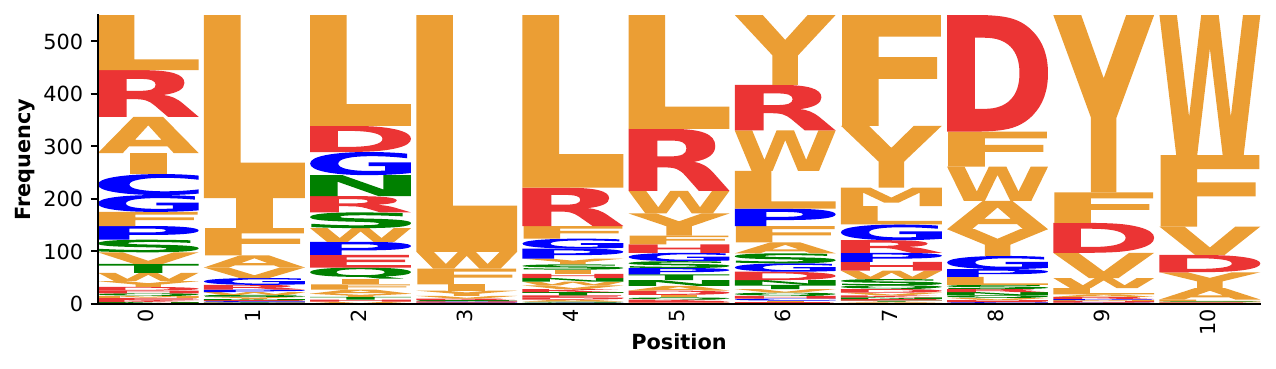}}
\end{center}
\end{minipage}
\begin{minipage}[t]{0.49\textwidth}
\begin{center}
\centerline{\includegraphics[width=1.0\textwidth]{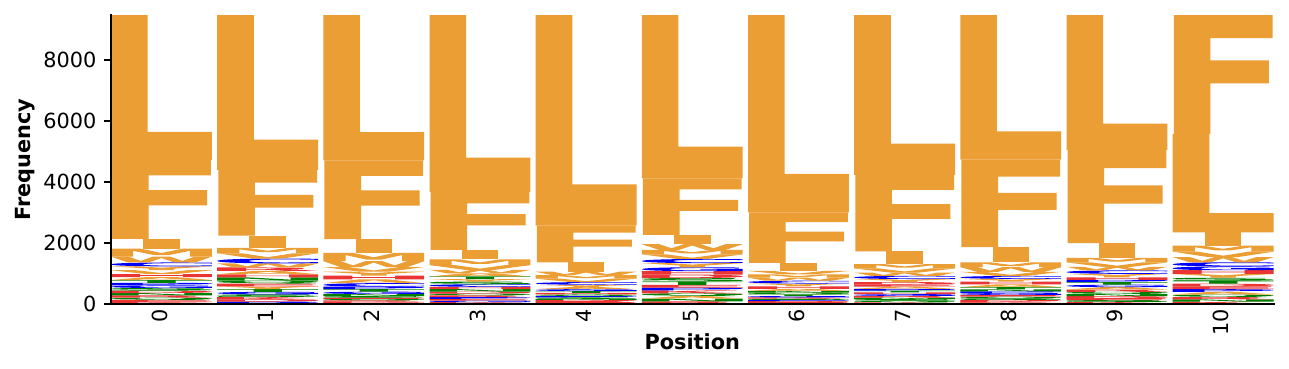}}
\end{center}
\end{minipage}

\caption{Amino acid frequency per position in the top 0.01\% CDRH3 sequences in the Absolut! database (left) and in CDRH3 sequences, reaching at least the same affinity, discovered in our experiments (right). Colors indicate the \gls*{aa} polarity.}
\label{logoplot}
\end{figure}

\paragraph{CDRH3 Analysis}
\Ours{} designs CDRH3 sequences with the sole objective of maximizing the affinity to a given antigen. Nonetheless, we analyze the \gls*{aa} distribution and binding specificity of designed sequences binding 1ADQ\_A to derive further insights. Thereby, we define a sequence as binding if it reaches the affinity of the top 1\% of 6.9 million murine CDRH3 sequences on the corresponding antigen in the \absolut database. We define a CDRH3 sequence as specific if it binds only one of the eight antigens ~\citep{zhou2009intrabody}. The logo plots shown in \figref{logoplot} illustrate the \gls*{aa} distributions of high-affinity CDRH3 sequences, revealing a bias toward leucine, phenylalanine, and more generally \glspl*{aa} with hydrophobic side chains. This coincides with the findings by \citet{cowen_rivers2022structured}. Additionally, we noted that the fraction of specific binders in the top 0.01\% CDRH3 sequences in the \absolut database is significantly higher than the fraction of specific binders, of equal or higher affinity, discovered in our experiments. We omit a more detailed analysis of the designed CDRH3 sequences regarding their biophysical properties as those are neither contained in \absolut software nor optimized by \ours{}. We suggest future work regarding antigen-specific CDRH3 design and the development of simulators incorporating more biophysical properties.

\paragraph{Offline Learning}
To evaluate our method's performance in the offline setting we copy the Replay and Fitness Buffers of the offline \gls*{rl} agents from a single online agent and train them without environment interaction. In \tabref{large-table} we present the binding energies reached by our offline agents as well as the best energy contained in their Replay Buffer (named Online Lead).
We observe that our agents can closely match this peak affinity and exhibit low standard deviations despite never interacting with the \gls*{mdp}. In \figref{progress_offline} we display the evaluation progress of our offline agents. It is evident that the majority of offline agents converge significantly faster and demonstrate similar or enhanced stability compared to their online counterparts. It is noteworthy that our method was not altered in any way to support offline learning. In \secref{stochastic} in the Appendix, we show preliminary results on how to use a trained offline agent to generate new CDRH3 sequences with improved energy distribution compared to the offline dataset.

\section{Conclusion}
We introduced a novel \gls*{rl} method that addresses the unique challenges of the antibody design task. Through experiments conducted on the \absolut software, we demonstrated its stability and effectiveness in designing high-affinity antibody CDRH3 sequences, surpassing the performance of existing methods. 
\Ours{} has the advantage of being applicable to both online and offline RL settings, enabling learning from pre-collected datasets. This capability has significant potential for real-world experiments since performing real-time evaluation of antibodies for online training is infeasible. Further work is recommended to expand \ours{} to include the design of antigen-specific CDRH3 sequences.

\section*{Acknowledgments}
This project was funded by the Mertelsmann Foundation. This work is part of BrainLinks-BrainTools which is funded by the Federal Ministry of Economics, Science and Arts of Baden-Württemberg within the sustainability program for projects of the excellence initiative II.
\bibliography{Stable_Online_and_Offline_Reinforcement_Learning_for_Antibody_CDRH3_Design}

\appendix
\section{Architecture}
\label{arch}
\begin{figure}[ht]

\begin{center}
\includegraphics[width=0.6\columnwidth]{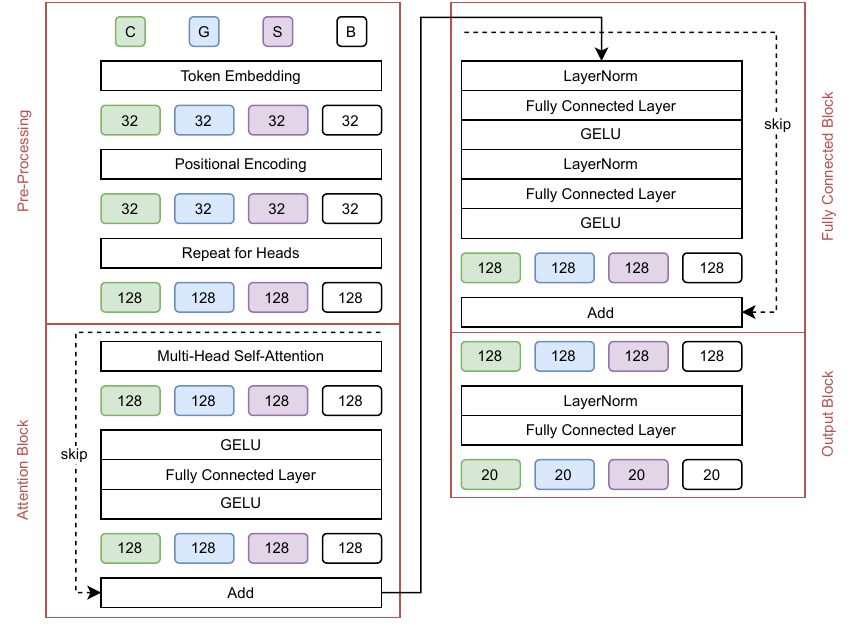}

\caption{Visual summary of our network architecture.}
\label{fig:arch}
\end{center}
\end{figure}

In the following, we give the details of our neural network architecture, which we visualize in \figref{fig:arch}.

On a high level, the architecture consists of pre-processing of the sequence, followed by an attention block, a fully connected block, and an output block. We add skip connections around the attention block and the fully connected block.

Pre-processing: We begin by embedding the discrete amino acid tokens into a learnable continuous 32-dimensional vector. We add a 32-dimensional learnable positional embedding to allow differentiating the same amino acid type at different positions in the sequence. We then repeat the embeddings four times, once per attention head.

Attention block: We apply LayerNorm, followed by multi-head self-attention using four heads. We add a look-ahead preventing mask to the attention logits before applying a softmax. This mask is implemented as a triangular matrix filled with $-\infty$. Therefore, the attention for tokens following the current token will become zero. After applying the attention to the value-heads output, we apply a GELU activation. As we utilize four heads, the resulting latent dimension is a 128-dimensional vector for each token in the sequence.
We project this latent representation using a fully connected layer, without LayerNorm as this would normalize the four different attention outputs, followed by another GELU activation. As we do not repeat this self-attention block, our architecture is not strictly a transformer but rather an attention-based fully connected network.

Fully connected block: We repeat two iterations of LayerNorm, fully connected layers, and GELU activation, without altering the latent dimension.

Output block: We apply a final LayerNorm before mapping the latent representation to a 20-dimensional output, representing Q-values for the twenty naturally occurring \glspl*{aa}.
To derive the \gls*{aa} to be placed at a specific token position, we evaluate the Q-values of the respective $B$ token. 

\section{Method Summary}
In \algref{alg:summary} we present an overview of our method. 
The fraction of samples from the Fitness Buffer is determined by the fitness fraction $f$.
Note, that this pseudocode depicts the online setting. For the offline setting lines 4-12 can be removed and $B$ and $F$ need to be initialized from a dataset.
\label{details}
\begin{algorithm}[ht]
   \caption{Stable Reinforcement Learning for Antibody Design}
   \label{alg:summary}
\begin{algorithmic}[1]
    \STATE {\bfseries Input:} ensemble size $N$, exploration probability $\epsilon$, reward scaling constant $\beta$, Fitness Buffer size $k$, replay ratio $RR$, batch size $b$, fitness fraction $f$, learning rate $\alpha$, target update ratio $\tau$,
   \STATE {\bfseries Initialize:} $N$ Q-networks $Q_{0,...,N-1}$ and target networks $Q'_{0,...,N-1}$, Replay Buffer $B$, Fitness Buffer $F$ of size $k$
   \FOR{$e=0$ {\bfseries to} $2499$}
   \STATE $s_0$= initial state of the MDP
   \FOR{$i=0$ {\bfseries to} $10$}
   \STATE $Q_{min}(s, a) \leftarrow \min_{i\in\{1,...,N\}} Q_i(s, a), \forall a \in A$
   \STATE get $\epsilon$-greedy action $a_i$ from $Q_{min}$
   \STATE Execute action $a_i$, receive $r_i$, $s_{i+1}$
   \STATE apply non-linear scaling $r_i\leftarrow(\frac{r_i}{\beta})^2$
   \ENDFOR
   \STATE Store trajectory $(s_i,a_i,r_i,s_{i+1}) \forall i\in [0,10]$ in $B$
   \STATE pushpop trajectory $(s_i,a_i,r_i,s_{i+1}) \forall i\in [0,10]$ onto $F$
   \FOR{0 {\bfseries to} $RR-1$}
   \STATE sample minibatch of size $b$ with transitions $(s_D,a_D,r_D,s'_D)$ from $B\cup F$ with $f\%$ from $F$
   \STATE $y = \max_{a'\in A}Q'_{min}(s'_D,a')$
   \STATE Update all Q-functions $Q_{i}(s_D, a_D)\leftarrow r_D+y$ using gradient descent with learning rate $\alpha$ 
   \ENDFOR
   \STATE Soft-update target network weights using $\tau$
   \ENDFOR
\end{algorithmic}
\end{algorithm}

\section{Hyperparameters}
\label{hyper}
Due to limited computational resources, we did not perform extensive hyperparameter sweeps but instead adapted the hyperparameters to optimize results in preliminary runs. We assume that \ours{} could benefit from a better choice of hyperparameters.

The ensemble size $N$ was set to 10 in the online setting while being set to 20 in the offline setting to achieve higher stability.
The remaining hyperparameters are shared for all experiments and are set as follows. Exploration probability ($\epsilon$) = 0.2, Reward scaling constant ($\beta$) = 70, Fitness Buffer size ($k$) = 10, Replay Ratio ($RR$) = 50, Batch size ($b$) = 64, Fitness fraction ($f$) = 20\%, Learning Rate ($\alpha$) = 0.0005, and Target Update Ratio ($\tau$) = 0.002.

\section{Statistical Significance}
\label{ttest-section}
To evaluate the significance of our improvements over other methods we utilize Welch's t-test on the final binding energy.
As in \secref{results}, the values for SQL were taken from the respective publication. However, the authors only provided integer-valued, presumably rounded results for some targets. 
The values for AntBO are derived from the published normalized values by scaling them with the normalization values given in the \absolut column of \tabref{large-table}.

P-values resulting from Welch's t-test give the probability that the null hypothesis, that is, the final energy scores of the two evaluated methods are drawn from equal distributions, holds true. We reject the null hypothesis for probabilities below 5\%.
When evaluating against \gls*{sql} we utilize the mean and standard deviations given over ten seeds and the mean and standard deviation of our method over eight seeds.
Note, that the presumably rounded values in the \gls*{sql} publication distort our analysis.
As no standard deviation for AntBO was published, we employ a one-sample t-test utilizing the provided mean value.
Given the available data, these tests allow, to the best of our knowledge, the least biased evaluation.

We find that \ours{} in the online setting gives a significant improvement compared to AntBO on all evaluated antigens. Despite reaching higher binding affinity than \gls*{sql} on all eight targets, the improvement is only significant on $\frac{3}{8}$ of the targets, namely 1ADQ\_A, 1NSN\_S, and 2JEL\_P. 
In the offline setting \ours{} again gives a significant improvement compared to AntBO on all eight antigens. Compared to \gls*{sql} we observe a significant improvement on all antigens except 2DD8\_S. 

\section{Ablation Study}
\label{ablation}
\begin{figure}[ht]

\begin{center}
\begin{minipage}[t]{0.49\textwidth}
\begin{center}
\centerline{\includegraphics[width=1.0\textwidth]{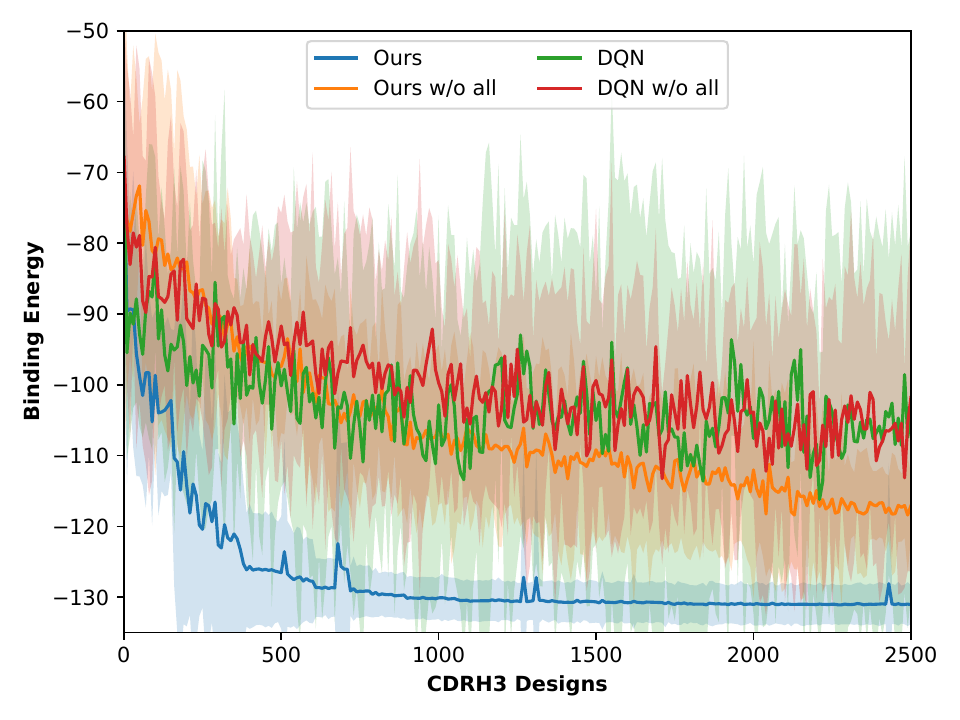}}
\end{center}
\end{minipage}
\begin{minipage}[t]{0.49\textwidth}
\begin{center}
\centerline{\includegraphics[width=1.0\textwidth]{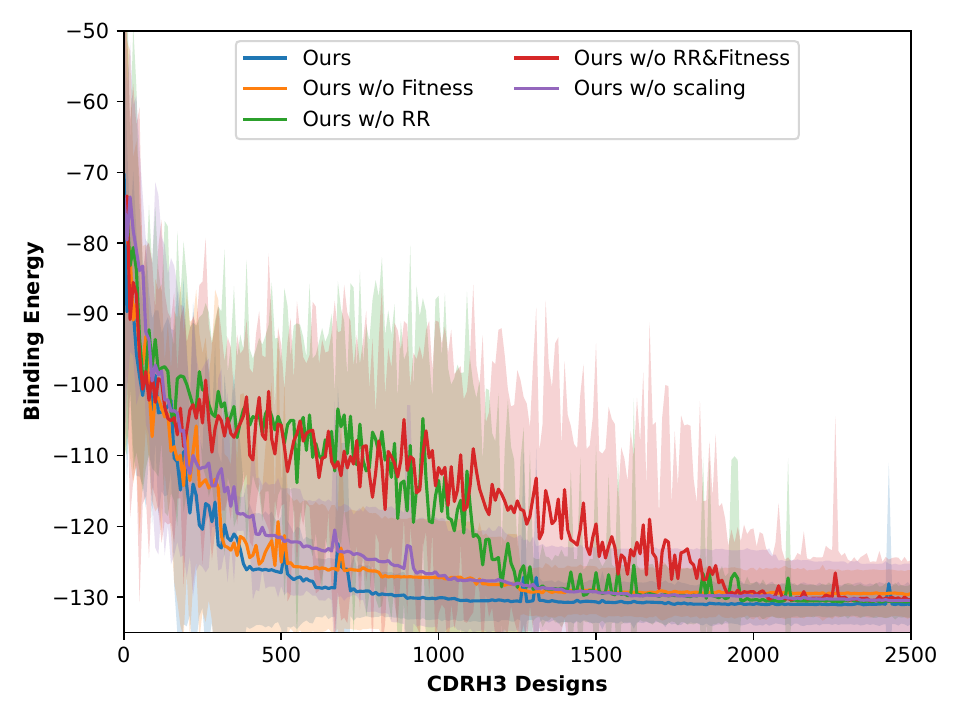}}
\end{center}
\end{minipage}
\caption{Progress of energy scores of our method and different ablation runs on the antigen 2DD8\_S. Mean and two standard deviations over eight seeds are displayed. ``Ours'' represents \ours{} with all its components. We refer to \ours{} without a high \gls*{rr}, Fitness Buffer, and Scaling as ``w/o all'' as that reflects \ours{} without all its components. On the left, we visualize the convergence of \ours{} in comparison to \gls*{dqn} both with and without all our components. On the right, we visualize the effect of removing a subsection of the components.}
\label{ablation-fig}
\end{center}
\end{figure}

To validate the effect of our chosen components we carried out multiple ablation studies on the antigen 2DD8\_S and further compared \ours{} to \gls*{dqn}~\citet{mnih2015human} with the same network architecture and hyperparameter setting.
We refer to \ours{} with all its components as ``Ours''. We evaluated removing all components (w/o all), the Fitness Buffer (w/o Fitness), reducing the \gls*{rr} to 10 (w/o RR), reducing the \gls*{rr} while removing the Fitness Buffer (w/o RR\&Fitness), and removing the nonlinear reward scaling (w/o Scaling). On the left side of \figref{ablation-fig}, we visualize the convergence of \ours{} in comparison to \gls*{dqn} both with and without all our components. On the right side, we visualize the effect of removing a subsection of the components.
We can observe that all components have an influence on the training progress, especially in the early training stage. Further, we see that DQN did not converge within 2500 training iterations, neither with nor without our introduced components. This shows the effect of the Maxmin-Ensemble, as \ours{} could be described as \gls*{dqn} utilizing said ensemble instead of a plain neural network.

\section{Stochastic CDRH3 Design}
\label{stochastic}
\begin{figure}[ht]
\begin{center}
\includegraphics[width=0.6\columnwidth]{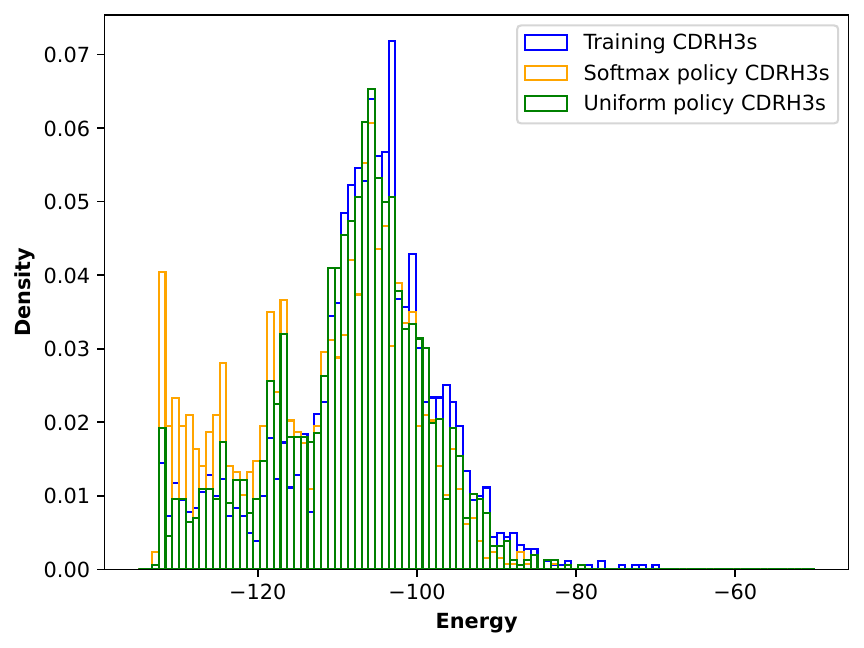}

\caption{Energy distributions over the offline training dataset as well as newly generated CDRH3 sequences using the softmax policy and uniform policy. Duplicated CDRH3 sequences have been removed.}
\label{fig:density}
\end{center}
\end{figure}

In this section, we present preliminary results.
We utilize a trained ensemble of Q-networks to generate novel CDRH3 sequences. In contrast to the deterministic Maxmin-policy we utilized in our online and offline experiments, we here define a stochastic policy based on a mixture of the deterministic $\argmax$ and a gumble-softmax distribution based on normalized Q-values which we refer to as the softmax policy.
This approach can theoretically be implemented with Q-networks from both online and offline settings and could also be used for exploration. Here, we test it with an ensemble of Q-networks trained offline on CDRH3 sequences explored on the antigen 2DD8\_S.
To verify that using the softmax policy  has a positive effect, we also added a control distribution which we generated by mixing the deterministic $\argmax$ and a uniform distribution over \glspl*{aa} called uniform-policy.
In \figref{fig:density} we visualize the density distributions over binding energies, showing an increase in high-affinity binders using the softmax sampling.
The distribution's mean and standard deviation shift from $-107.04\pm 9.85$ to $-108.55\pm9.55$ and $-11.76\pm10.75$ using the uniform and softmax policy respectively. All those shifts are significant with p-values below $1\mathrm{e}{-5}$. Using the softmax policy we further generated multiple new best binders outside the training distribution. Note, that this distribution is sensible to the fraction of \glspl*{aa} derived using the $\argmax$. A higher fraction leads to higher average affinity but also more duplicated CDRH3 sequences in the set of newly generated sequences. 
Further, we found that out of $\sim1500$ unique CDRH3 sequences generated using this approach, only $\sim100$ were contained in the training distribution, showing that this approach can be used to generate new CDRH3 distributions with improved energy distributions.

\end{document}